\documentclass[twocolumn,showpacs,showkeys,superscriptaddress]{revtex4}
\usepackage{amsmath}
\usepackage{amsfonts}
\usepackage{amssymb}
\usepackage{color}
\usepackage{dsfont}

\begin{document}

\title{Correspondence between dark energy quantum cosmology and Maxwell equations}
\author{Felipe A. Asenjo}
\email{felipe.asenjo@uai.cl}
\affiliation{Facultad de Ingenier\'{\i}a y Ciencias,
Universidad Adolfo Ib\'a\~nez, Santiago 7941169, Chile.}

\author{Sergio A. Hojman}
\email{sergio.hojman@uai.cl}
\affiliation{Departamento de Ciencias, Facultad de Artes Liberales,}
\affiliation{Departamento de F\'{\i}sica, Facultad de Ciencias, Universidad de Chile,
Santiago, Chile.}
\affiliation{Centro de Recursos Educativos Avanzados,
CREA, Santiago, Chile.}

\begin{abstract}
 A Friedmann-Robertson-Walker cosmology with dark energy can be modelled using a quintessence field. That system is equivalent to a relativistic particle moving on a two-dimensional conformal spacetime. When the  quintessence behaves as a  free massless scalar field in a Universe with cosmological constant, the quantized version of that theory can lead to a Supersymmetric Majorana quantum cosmology. The purpose of this work is to show that such quantum cosmological model correspond to the Maxwell equations for electromagnetic waves propagating in a medium with specific values for its relative permittivity and relative permeability. The  form of those media parameters are calculated, implying that a Majorana quantum cosmology can be studied in an analogue electromagnetic system.
\end{abstract}


\maketitle

\section{Introduction}

The aim of this work is to show that there exists a correspondence between the seemingly different physical models of a cosmological model using dark energy and Maxwell equations.
The link between these two formalisms appears when one considers the quantum version of the cosmological model  \cite{SupMajQCos} using the Breit prescription \cite{breit} for spin particles.

The representation of dark energy using a quintessence field with a potential, allows us to describe the cosmological dynamics in  fashion which is analogous to the description of the dynamics of a relativistic particle. In the case of quintessence described by a free massless scalar field in a Universe with cosmological constant, the resultant theory  may be quantized by using a Klein-Gordon scheme, giving rise to the Wheeler-DeWitt equation \cite{dewitt,hawk1,alvarenga, oliveraG, lemos, mone, mone2, vaki, huang, barbour,hawk2}. However, using the Breit prescription \cite{breit}, the same model can be quantized as a spinorial theory. This procedure yields a Majorana version for the Quantum cosmology which happens to be supersymmetric \cite{SupMajQCos}.

In addition of the  cosmological implications of such theory, the aim of this article is to show its direct correspondence with the description of  propagating electromagnetic fields in a medium using Maxwell equations. We can identify the relative permittivity and permeability of the medium with parameters of the quantum cosmological model. As we show, this implies that the Supersymmetric Majorana quantum cosmology can be  studied in an analogue electromagnetic system using either
normal materials or  negative-index metamaterials (NIMs) \cite{engheta,Eleftheriades}.

\section{Quantum cosmology with dark energy}

Consider  an isotropic and homogeneous Friedmann-Robertson-Walker (FRW) spacetime with a line element \cite{ryden}
\begin{equation}
ds^2=dt^2- a(t)^2\left[\frac{dr^2}{1-kr^2} +r^2(d\theta^2+
\sin^2\theta d\phi^2)\right]\, ,
\end{equation}
where $a(t)$ is the scale factor, and the curvature constant $k=\pm 1,0$.
The evolution of a FRW cosmology with cosmological constant $\Lambda$, interacting with a quintessence (massless scalar) field $\phi(x^\beta)$ characterized by a potential ${\cal V}(\phi)$, can be found by using Einstein equations \cite{SupMajQCos} ($8\pi G/c^4=1$, where $G$ is the gravitational constant and $c$ is the speed of light)
\begin{eqnarray}
2\frac{\ddot a}{a}+\left(\frac{\dot a}{a}\right)^2 +\
\frac{k}{a^2} +\frac{1}{2}\ \dot \phi^2-V(\phi)&=& 0\, ,\label{frwqe1} \\
3\left(\frac{\dot a}{a}\right)^2 +3\frac{k}{a^2}-\left(\frac{1}{2}\dot \phi^2+ V(\phi) \right) &=&0\, , \label{frwqc}
\end{eqnarray}
where we introduced $V(\phi)={\cal V}(\phi) - \Lambda$.
Also,
the Klein-Gordon equation for the quintessence field is
\begin{equation}
\ddot\phi + 3\frac{\dot a}{a} \dot\phi + \frac{d
V(\phi)}{d\phi}=0\, . \label{frwqe2}
\end{equation}
The above system
describes the evolution of a Friedmann--Robertson--Walker--Quintessence (FRWQ) Universe.

In Ref.~\cite{SupMajQCos} was shown that the Lagrangian
\begin{equation}
L_F=  \sqrt{\bar V(\xi, \theta)\,  e^{2\xi}\left( \dot\theta^2- \dot\xi^2\right)}\, ,\label{LF2}
\end{equation}
 gives rise to all three equations
\eqref{frwqe1},  \eqref{frwqc} and \eqref{frwqe2}, with variables $\xi=\ln({2\sqrt 6} a^{3/2}/3)$,  $\theta={3\phi}/({2\sqrt 6})$ [here $\dot\xi={d\xi}/{d\lambda}$ and  $\dot\theta={d\theta}/{d\lambda}$, with $\lambda$ is an arbitrary parameter], and  the general  potential
\begin{equation}
\bar V(\xi,\theta)=3 \left(\frac{3}{8}\right)^{{1}/{3}}k\, e^{2\xi/3}-\left(\frac{3}{8}\right) e^{2\xi} V(\theta)\, .
\end{equation}
This Lagrangian  \eqref{LF2}  shows  that the FRWQ cosmology evolves as a relativistic particle moving in two dimensional spacetime under the influence of potential $\bar V$.
It can be proved that the  Lagrangian  \eqref{LF2}  gives rise to the FRWQ equations by recalling that the Jacobi--Maupertuis and Fermat principles  \cite{h2} yield
identical equations of motion in classical mechanics and geometrical
(ray) optics except for the fact that Fermat principle also produces
a constraint equation. 
Notice that Lagrangian \eqref{LF2} can be written as the one for a relativistic particle in a two-dimensional conformally flat spacetime 
\begin{equation}	
L_F=  \sqrt{g_{\mu\nu} \left(\frac{dx^\mu}{d\lambda}\right) \left(\frac{dx^\nu}{d\lambda}\right)}\, ,
\end{equation}
with the metric $g_{\mu\nu}=\Omega^2\, \eta_{\mu\nu}$ (where $\eta_{\mu\nu}$ is the flat spacetime metric, and $x^0=\theta$ and $x^1=\xi$), and the conformal factor 
\begin{equation}\label{functionOmegaa}
\Omega\equiv\sqrt{\bar V}e^\xi=\left[3 \left(\frac{3}{8}\right)^{\frac{1}{3}}k\, e^{8\xi/3}
-\frac{3}{8} V(\theta) e^{4\xi}\right]^{1/2}\, ,
\end{equation}
Thus, the FRWQ system is equivalent to a relativistic particle moving in a two-dimensional conformally flat spacetime, where the quintessence field plays the role of an effective time.

The above results allow to find the quantum version of the FRWQ cosmology, through quantization of the Lagrangian \eqref{LF2}.
 We restrict ourselves to cases where $\bar V >0$, in order to avoid problems in the quantization procedure. 
We also
 consider static manifolds  \cite{saa} such that there exists a family of spacelike surfaces orthogonal to a timelike Killing vector. 
Therefore, Lagrangian \eqref{LF2} is always real, which can be proved by using the definitions of $\xi$, $\theta$ and $\bar V$.
All of this implies that ${\partial_\theta g_{\mu\nu}}=0$, or ${\partial_\theta}\bar V=0$,  describing a constant potential $V(\theta)$ (becoming essentially equal to a cosmological constant).
Thus, the associated Noether conservation law implies that $\dot\theta$ does not change
sign,  and  $\theta$ can be used as the evolution time variable.
Using this $\theta$--time, it can be  shown at classical level \cite{hanson} that the Hamiltonian for  Lagrangian \eqref{LF2} is
$  H=\sqrt{g_{00}}\sqrt{1-g^{11}\pi^2}=\sqrt{g_{00}}\sqrt{1+{\pi^2}/{\Omega^{2}}}$,
where $\pi$ is the canonical momentum, and $\sqrt{g_{00}}=\Omega$.
This classical Hamiltonian is used to construct the  quantum Hamiltonian operator $\mathcal H$ as \cite{SupMajQCos}
\begin{equation}\label{momentumoperator1}
  \mathcal H=\Omega^{1/2}\sqrt{1+\hat p^2}\, \Omega^{1/2}\, , \quad \hat p=\sqrt{-g^{11}}\hat\pi=-\frac{i}{\Omega}\frac{\partial}{\partial\xi}\, ,
\end{equation}
where $\hat p$ is the momentum operator  and $\hat\pi=-i{\partial}_\xi$.
 Therefore, the quantum equation that describes the quantization of the FRWQ system is
\begin{equation}\label{quantumequtiona}
i\hbar\frac{\partial\Psi}{\partial\theta}=\mathcal H\Psi\, ,
\end{equation}
where $\Psi$ is the wavefunction for the quantum FRWQ cosmology.
The quantum equation \eqref{quantumequtiona} emerges because of the direct correspondence between the FRW geometry and the quintessence scalar field at a Fermat-like Lagrangian level. To find a solution, we need  a quantization procedure to solve  the square-root of the Hamiltonian operator \eqref{momentumoperator1}. It is costumary to solve  the square-root using a spinless particle approach to get a Klein-Gordon equation \cite{gavrilov}, which gives origin to the Wheeler--DeWitt Super-Hamiltonian formalism.
However, notice that there is no restriction for the quantization scheme possible to be used.
In principle, we can solve the square-root using matrices, obtaining the quantization of a relativistic particle which leads to the Dirac equation. This procedure was introduced by Breit \cite{breit}, that  shows that there is a correspondence between the Dirac and the relativistic pointlike particle Hamiltonians.
Breit's interpretation  \cite{breit}  identifies the Dirac matrices as ${\pi}/{H}\rightarrow \alpha$, and $\sqrt{1-({\pi}/{H})^2}\rightarrow\beta$.  These identifications are consistent with the postulates of Dirac electron's theory \cite{breit,savasta}.
By the Breit's prescription, the Hamiltonian operator \eqref{momentumoperator1} becomes \cite{SupMajQCos}
\begin{equation}\label{momentumoperatorDirac}
  {\cal H}=\Omega^{1/2} \left(\alpha\cdot\hat p+\beta\right) \Omega^{1/2}\, ,
\end{equation}
where $\alpha$ and $\beta$ are the two-dimensional flat spacetime Dirac matrices, as the effective curvature is already taken into account in $\Omega$ (with $\hbar=1$).
Moreover, now the wavefunction $\Psi$ [in Eq.~\eqref{quantumequtiona}]
is a two-dimensional spinor.

With this Hamiltonian, and defining the wavefunction $\Phi=\sqrt{\Omega}\Psi$, we finally find from  Eq.~\eqref{quantumequtiona}  the  spinor quantum equation \cite{SupMajQCos,symmetryQcosmoCAH}
\begin{equation}\label{diracQQQ2}
 i\gamma^0\frac{\partial\Phi}{\partial\theta}+  i\gamma^1\frac{\partial\Phi}{\partial\xi}=\Omega\Phi\, .
\end{equation}
where
$\gamma^0=\beta$ and $\gamma^1=\gamma^0\alpha$.  The above equation corresponds to a Quantum Cosmology theory for the FRWQ system, modelling now the Universe as a spin particle in a two-dimensional conformally flat spacetime \cite{SupMajQCos}.

In order to obtain real wavefunctions $\Phi$, the matrices in Eq.~\eqref{diracQQQ2} should correspond to the two-dimensional Majorana representation
\begin{equation}\label{MajoranaMatrices}
\gamma^0=\left(\begin{array}{cc}
                          0 & -i\\
                          i &0\\
                        \end{array}
                      \right)\, , \qquad
\gamma^1=\left(\begin{array}{cc}
                          i & 0\\
                          0 &-i\\
                        \end{array}
                      \right)\, .
\end{equation}
In this form,
 Eq.~\eqref{diracQQQ2} becomes a set of supersymmetric equations of quantum mechanics \cite{death,vargas,cooper,crom,cooper2,SupMajQCos}, which can only be obtained in the Majorana picture.
The implications of this system were thoroughly studied in Ref.~\cite{SupMajQCos}, showing that Eq.~\eqref{diracQQQ2} with matrices \eqref{MajoranaMatrices} represents a Supersymmetric Majorana quantum cosmology.
This can be seen by defining
\begin{equation}
\Phi(\theta,\xi)=\left(\begin{array}{c}
                          \varphi_+(\xi) e^{E\theta}\\
                          \varphi_-(\xi) e^{-E\theta} \\
                        \end{array}
                      \right)\, , 
\end{equation}
to find from \eqref{diracQQQ2} the set of supersymmetric equations of quantum mechanics \cite{SupMajQCos,death,vargas,cooper,crom,cooper2,symmetryQcosmoCAH}
\begin{equation}\label{SUSYM}
Q_\pm \varphi_\pm=E\varphi_\mp\, .
\end{equation}
with the operators $Q_\pm=\pm{d_\xi}+\Omega$.
These equations are supersymmetric with the two spinor components being super--partners of each other.
 Each wavefunction satisfies $H_\pm\varphi_\pm={E^2}\varphi_\pm$, with the Hamiltonians operators $H_\pm=-{d_\xi^2}+{\cal W}_\pm$, and potentials ${\cal W}_\pm=\mp{d_\xi\Omega}+\Omega^2$ \cite{SupMajQCos}. Also, $Q_\pm$ correspond to supercharge operators. This theory can produce new versions of quantum cosmological models. For example, it can be shown that in flat curvature case, the Universe behaves as a diatomic molecule subject to the Morse potential \cite{SupMajQCos}.

\section{Correspondence to Maxwell equations}

In this section we focus the correspondence between the theory  \eqref{diracQQQ2} and  electromagnetism.
In order to make this correspondence manifest, let us consider Maxwell equations in a medium  with permittivity $\epsilon$ and permeability $\mu$ in the absence of charges \cite{jackson}
\begin{eqnarray}\label{maxweltotales}
\frac{\partial{\bf D}}{\partial t}&=&\nabla\times{\bf H}\, ,\nonumber\\
\frac{\partial {\bf B}}{\partial t}&=&-\nabla\times{\bf E}\, ,
\end{eqnarray}
with the electric field ${\bf E}$, the magnetic field ${\bf B}$, the displacement field ${\bf D}=\epsilon\, {\bf E}$,  the magnetization field ${\bf H}={\bf B}/\mu$ (chosing the speed of light $c=1$). 
Furthermore, $\nabla\cdot{\bf D}=0$ and $\nabla\cdot{\bf B}=0$.

In order to show the correspondence with the Quantum Cosmology model, we assume that the permittivity  and the permeability are not constant. Also, let us consider a two-dimensional spacetime system (one temporal and one spatial dimension), with spatial variations in, let us say, the $\hat e_z$-direction. We choose transverse fields ${\bf B}(t,z)=B(t,z) \hat e_x$, and  ${\bf D}(t,z)=D(t,z) \hat e_y$, such that ${\bf B}\cdot\hat e_z=0={\bf D}\cdot\hat e_z$, and ${\bf B}\cdot{\bf D} =0$.
Besides, the time-independent permittivity  and permeability have the same spatial dependence, $\epsilon=\epsilon(z)$ and $\mu=\mu(z)$. Then, Maxwell equations \eqref{maxweltotales} acquire the form
\begin{eqnarray}\label{ecMaxwe12ws}
\frac{\partial {\cal D}}{\partial t}&=&-\frac{B}{\hat\mu^2}\frac{\partial\hat\mu}{\partial z}+\frac{1}{\hat\mu}\frac{\partial B}{\partial z}\, ,\nonumber\\
\frac{\partial B}{\partial t}&=&-\frac{{\cal D}}{\hat\epsilon^2}\frac{\partial \hat\epsilon}{\partial z}+\frac{1}{\hat\epsilon}\frac{\partial {\cal D}}{\partial z}\, .
\end{eqnarray}
where ${\cal D}=\mu_0 D$, the relative permittivity is $\hat\epsilon=\epsilon/\epsilon_0$, the relative permeability is $\hat\mu=\mu/\mu_0$, and $\epsilon_0$ and $\mu_0$ are the free-space permittivity  and permeability respectively (with $\epsilon_0\mu_0=1$).

By introducing the spinor
\begin{equation}\label{spinorEM}
\Phi=\left(\begin{array}{c}
                          B \\
                          {\cal D}
                        \end{array}
                      \right)\, ,
\end{equation}
we can rewrite system \eqref{ecMaxwe12ws} in a simple way
as
\begin{equation}\label{ecMaxwellDira}
i\gamma^0\frac{\partial \Phi}{\partial  t}+i\Gamma_1\frac{\partial \Phi}{\partial z}=\Gamma_2\Phi\, ,
\end{equation}
with $\gamma^0$ given in \eqref{MajoranaMatrices}, and the matrices
\begin{equation}\label{matricesam12}
\Gamma_1=\left(\begin{array}{cc}
                          i/\hat\mu & 0 \\
                          0 & -i/\hat\epsilon
                        \end{array}
                      \right)\, ,\quad
\Gamma_2=\left(\begin{array}{cc}
                          -\hat\mu'/{\hat\mu^2} & 0 \\
                          0 & \hat\epsilon'/\hat\epsilon^2
                        \end{array}
                      \right)\, ,\quad
\end{equation}
with $\hat\mu'={\partial \hat\mu}/{\partial z}$ and $\hat\epsilon'={\partial \hat\epsilon}/{\partial z}$.
 The spinor form \eqref{ecMaxwellDira} of Maxwell equations  is general for a two-dimensional spacetime. For any
general permittivity and permeability, Eq.~\eqref{ecMaxwellDira} does not coincide with the Quantum cosmology equation \eqref{diracQQQ2}. For example, for vacuum, $\Gamma_2=0$.

However, we can show that there exists a regime in which Maxwell equations and the  Supersymmetric Majorana Quantum Cosmology coincide. Let us consider the following form for the relative permittivity and relative permeability
\begin{eqnarray}\label{permittivitypermeabilityQ}
\hat\epsilon(z)\approx 1+\lambda(z)\, ,\quad \hat\mu(z)\approx 1-\lambda(z)\, ,
\end{eqnarray}
in terms of a function $\lambda$ to be determined.
We focus our attention in the regime when  $|\lambda|\ll 1$.
The relative permeability and permittivity of this material are almost equal, $\sqrt{\hat\mu/\hat\epsilon} \approx 1-\lambda$. With the relative
permittivity and permeability given by \eqref{permittivitypermeabilityQ}, the  matrices \eqref{matricesam12} becomes
\begin{eqnarray}\label{matricesam12aaa}
\Gamma_1&\approx&\left({1+\lambda^2}\right) \gamma^1+{i\lambda}\,  \mbox{\bf{1}} \, ,\nonumber\\
\Gamma_2&\approx&\frac{d \lambda}{d z}\,  \mbox{\bf{1}}-i\frac{d\lambda^2}{d z}\gamma^1\, ,
\end{eqnarray}
with the matrix $\gamma^1$ given by \eqref{MajoranaMatrices}, and the two-dimensional unit matrix ${\bf 1}$.
In general, for any $\hat\epsilon$ and $\hat\mu$, Eq.~\eqref{ecMaxwellDira}  does not have a conserved current $\bar\Phi\gamma^\mu\Phi=\Phi^\dag\Phi+\Phi^\dag\gamma^0\gamma^1\Phi$. Nonetheless, for materials satisfying 
\eqref{permittivitypermeabilityQ} and \eqref{matricesam12aaa}, Eq.~\eqref{ecMaxwellDira} can have a conserved current if $|\lambda|\ll 1$ and  that the variation ranges of $B$ and $D$ are much larger than the variation range of $\lambda$, i.e., $\lambda(\partial\Phi/\partial z)\ll(d\lambda/dz)\Phi$. Using these approximations, Maxwell equation \eqref{ecMaxwellDira} can be written in an approximated form as
\begin{equation}\label{ecMaxwellDira5}
i\gamma^0\frac{\partial \Phi}{\partial t}+i\gamma^1\frac{\partial \Phi}{\partial z}=\frac{d \lambda}{d z}\Phi\, ,
\end{equation}
which has the usual conservartion law for a Dirac equation.
With all of the above, the correspondence between Maxwell equations \eqref{ecMaxwellDira5} and
the  Supersymmetric Quantum Majorana Cosmologies equation \eqref{diracQQQ2} is now evident by  redefining
\begin{eqnarray}\label{resultcoreespondo}
t&=&\theta\, ,\nonumber\\
\xi&=& z\, ,\nonumber\\
\lambda&=&\lambda(\xi)= \int \Omega\,  d\xi=-\frac{4}{3V}e^{-4\xi}\, \Omega^3\, .
\end{eqnarray}
This last equation implies that this correspondence is only valid in the regime ${\Omega^3}e^{-4\xi}\ll {|V|}$, which is the limit needed in order to keep current conservation.

Results \eqref{resultcoreespondo} establish the complete correspondence between a Supersymmetric Majorana Quantum cosmological model and Maxwel equations. 
Materials satisfying \eqref{permittivitypermeabilityQ} must have almost equal  relative permeability and permittivity, and they can be both positive or both negative. In the former case, we are describing  a normal material with such a property. In case that both permeability and permittivity are negative, we are describing  a NIM \cite{engheta,Eleftheriades,Kraftmakher}. 
In general, the refraction index is
\begin{equation}
n(\xi)=\sqrt{\hat\epsilon(\xi)\hat\mu(\xi)}\approx \pm \left(1-\frac{8}{9V^2}e^{-8\xi} \Omega^6\right)\, ,
\end{equation}
where the positive sign describes a normal material, and the negative refractive index represents a NIM in which a wave propagates backwards.

There are materials with relative permittivity and relative permeability that can have the form \eqref{permittivitypermeabilityQ}.
Composite ferrites \cite{Zheng,Su,Kong,Thakur} can, under appropriated conditions, achieve almost matching permeability and permittivity values by shining radiation of different frequencies on the material.
This implies that the above results  can be tested in an analogue fashion using those materials. For example, for a spatially flat cosmology, the Maxwell equations and the Supersymmetric Majorana quantum cosmology coincide for $\lambda(\xi)=\sqrt{-3 {V}/32} \exp(2\xi)$, with constant $V<0$, and refraction index $n(\xi)\approx\pm [1+3 V\exp(4\xi)/64]$.
For this case, Maxwell equations can only describe a quantum cosmology for $\xi\ll \ln(11/|V|)$. Furthermore, the boundary conditions for $\Phi$ given in Eq.~\eqref{spinorEM} (which depend on $\Omega$) should be suitable for simulating a quantum cosmology. 
 For $k=0$, the boundary conditions for $\xi\rightarrow \infty$ ($a\rightarrow  \infty$) can be established to obtain a vanishing magnetic field at infinity in one spatial dimension (see Ref.~\cite{SupMajQCos}).

Our proposal is in the same spirit than similar ones for analogue optical systems for quantum cosmologies \cite{Westerberg,Batista}, and for gravity in general (see for example Refs.~\cite{VisserCarlos,VisserCarlos2,Ornigotti,Faccio}).
However, our result establishes an ``optical" analogue for a new kind of spinor quantized cosmological model. The proposed analogue electromagnetic media that correspond to the quantum cosmology
 is time-independent but space-dependent, which is an approach opposite to previous attempts \cite{Westerberg}. Relations \eqref{permittivitypermeabilityQ} are satisfied by certain tunable metamaterials \cite{Bi,grant}
and composite ferrites \cite{Zheng,Su,Kong,Thakur} used to operate at a wider range of frequencies.
All of the above makes of this Majorana Supersymetric quantum cosmological model a system worth to be studied by studying wave propagation in Maxwell equations in the appropriated media.

\end{document}